\documentclass[aps,pra,twocolumn,showpacs,amssymb,amsmath,amsfonts,floatfix,nofootinbib,groupedaddress,notitlepage]{revtex4}
\usepackage{graphicx,graphics,color,times,bm,bbm,hyperref}

\let\oldsqrt\sqrt
\def\sqrt{\mathpalette\DHLhksqrt}
\def\DHLhksqrt#1#2{%
\setbox0=\hbox{$#1\oldsqrt{#2\,}$}\dimen0=\ht0
\advance\dimen0-0.2\ht0
\setbox2=\hbox{\vrule height\ht0 depth -\dimen0}%
{\box0\lower0.4pt\box2}}

\begin{document}

\title{Distillation of free entanglement from bound-entangled states using weak measurements}

\author{S. Baghbanzadeh}
\affiliation{Department of Physics, Sharif University of Technology, Tehran, Iran}

\author{A. T. Rezakhani}
\affiliation{Department of Physics, Sharif University of Technology, Tehran, Iran}

\begin{abstract}
We propose a scheme for distillation of free bipartite entanglement from bipartite bound-entangled states. The crucial element of our scheme is an ancillary system that is coupled to the initial bound-entangled state via appropriate weak measurements. We show that in this protocol free entanglement can be always generated with nonzero probability by using a single copy of the bound-entangled state. We also derive a lower bound on the entanglement cost of the protocol and conclude that, on average, applying weaker measurements results in relatively higher values of free entanglement as well as lower costs.
\end{abstract}
\pacs{03.67.-a, 03.65.Ud, 03.65.Ta}
\maketitle

\section{Introduction}

Entanglement~\cite{QuantEnt} is a physical resource that plays a leading role in performing quantum computation and quantum information processing~\cite{Nielsen:book}. This is also helpful in understanding relevant properties of many-body quantum systems. It has been shown~\cite{Osterloh,wsls} that a singularity in the entanglement profile of the ground state of a many-body system, even a change in the type of its entanglement from bound (free) to free (bound)~\cite{BEQPT}, can be accompanied by quantum phase transitions.

Bound-entangled states are states from which no pure entanglement can be extracted only by local operations and classical communications (LOCC)~\cite{BE,Alber}. However, recent studies unveil the usefulness of these states in some protocols such as secure quantum key distribution~\cite{Hor-key}, remote quantum information concentration~\cite{remote}, quantum data hiding~\cite{hiding}, channel discrimination~\cite{chdis}, and reducing communication complexity~\cite{compelexity}. It was also indicated~\cite{act,Masanes} that application of any bound-entangled state along with some free-entangled state can increase the teleportation power of the free-entangled state. Preparation of some particular bipartite and multipartite bound-entangled states is now possible in nuclear magnetic resonance~\cite{NMR}, optical~\cite{Smolin1,several,unconditional,BeW,deGaussification}, as well as ion-trapped systems~\cite{decoherence}. Moreover, bound entanglement can naturally arise, e.g., in the XY spin model through applying an external magnetic field~\cite{magneticB}, in the Jaynes-Cummings model~\cite{JC}, as well in strongly-correlated graph states at thermal equilibrium~\cite{graph1,graph2}.

Bound entanglement can be ``activated" (or ``unlocked") into free entanglement~\cite{QuantEnt,act}. For example, in the multipartite case, a distillable ensemble can be obtained from a tensor product or a mixture of some non-distillable (i.e., bound) ensembles~\cite{superactivation,BI}. Additionally, in multipartite bound-entangled states, if two parties carry out a Bell-type measurement on their particles, the other parties can distill free entanglement by LOCC~\cite{Smolin,DC}. Alternatively, in the bipartite case, both unitary~\cite{unitary} and nonunitary~\cite{non-unitary} evolutions of bound-entangled states may result in the birth of free entanglement.

Here, we demonstrate that, by attaching an ancillary qubit to a bipartite bound-entangled state and performing two weak measurements between each party of that state and the ancilla, one can transform this bound entanglement to free entanglement with a nonvanishing probability. Measurements we employ here are ``weak'' in that they do not disturb the initial state strongly; thus, after performing the weak measurements, there is some probability with which strong or projective measurement does not occur~\cite{WM1,WM2,Igain}. Note that we do not rule out the existence of similar distillation scenarios with strong or projective measurements; here we only focus on a weak-measurement scenario. Our approach has a lower cost than the protocols suggested in Refs.~\cite{Smolin,DC,exp1} because, (i) the measurement operators we use have less nonlocal content than the Bell-type measurements, and (ii) we do not need to share any free-entangled state between parties. Furthermore, unlike existing distillation~\cite{distill1,distill2,distill3} and entanglement generation~\cite{entgen} protocols, we do not use any unitary operation either. Some environmentally-induced decoherence scheme may inhibit applying unitary operations, whence our scheme provides a controllable method for distillation of free entanglement from bipartite bound-entangled states.

\section{Bound-entangled states and weak measurement setting}

We illustrate our scheme through three qutrit-qutrit bound-entangled states shared between an ``Alice'' and a ``Bob'': a state which is complementary to the tiles unextendible product basis~\cite{UPB}; $\chi_1=\frac{1}{4}(\openone-\sum_{i=0}^4 |\psi_i\rangle\langle\psi_i|)$, in which
$|\psi_0\rangle=|0\rangle(|0\rangle-|1\rangle)/\sqrt{2}$, $|\psi_1\rangle=(|0\rangle-|1\rangle)|2\rangle/\sqrt{2}$, $|\psi_2\rangle=|2\rangle(|1\rangle-|2\rangle)/\sqrt{2}$, $|\psi_3\rangle=(|1\rangle-|2\rangle)|0\rangle/\sqrt{2}$, and $|\psi_4\rangle=(|0\rangle+|1\rangle+|2\rangle)(|0\rangle+|1\rangle+|2\rangle)/3$; and two Horodeckis' states~\cite{BEH}
\begin{eqnarray}
& \chi_2(a)= \frac{1}{1+8a}\left(
  \begin{array}{ccccccccc}
    a & 0 & 0 & 0 & a & 0 & 0 & 0 & a \\
    0 & a & 0 & 0 & 0 & 0 & 0 & 0 & 0 \\
    0 & 0 & a & 0 & 0 & 0 & 0 & 0 & 0 \\
    0 & 0 & 0 & a & 0 & 0 & 0 & 0 & 0 \\
    a & 0 & 0 & 0 & a & 0 & 0 & 0 & a \\
    0 & 0 & 0 & 0 & 0 & a & 0 & 0 & 0 \\
    0 & 0 & 0 & 0 & 0 & 0 & \frac{1+a}{2} & 0 & \frac{\sqrt{1-a^2}}{2} \\
    0 & 0 & 0 & 0 & 0 & 0 & 0 & a & 0 \\
    a & 0 & 0 & 0 & a & 0 & \frac{\sqrt{1-a^2}}{2} & 0 & \frac{1+a}{2} \\
  \end{array}
\right),\nonumber
\end{eqnarray}
\begin{figure}[tbp]
    \includegraphics[scale=.76]{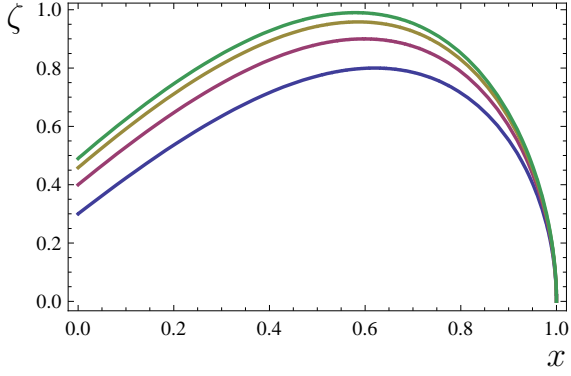}
    \caption{(Color online) Degree of weakness of the measurement, $\zeta$, vs parameter $x$ for $\beta=1/10,\;1/5,\;3/10$, and $2/5$ (from bottom to top). $\zeta$ is symmetric around $\beta=1/2$. According to the argument bellow Eq.~(\ref{weakness}), it is clear that for all $\beta$s, the measurement is strong when $x=1$. Depending on the value of $\beta$, the measurement is significantly weak when $x\in(1/\sqrt{3},1/\sqrt{2})$.}
    \label{fig1}
\end{figure}
\begin{figure*}[tbp]
      \includegraphics[scale=.74]{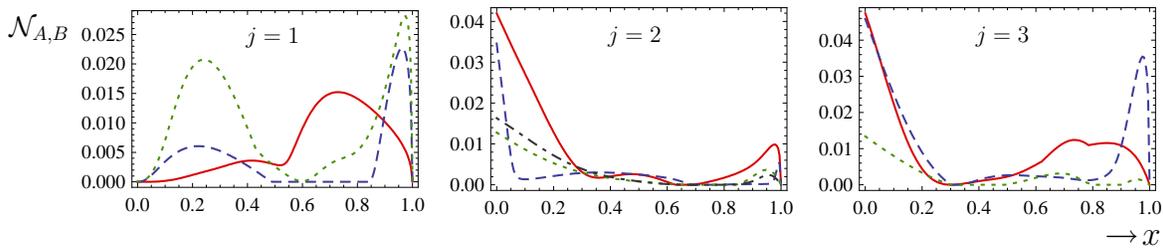}
    \caption{(Color online) Free entanglement between Alice and Bob for different outcomes $i=1$, $2$ (dashed), and $3$ (dotted and dot-dashed curve in the middle panel) in terms of $x$ when the initial state is $\chi_1$. Except the dot-dashed curve (which corresponds to the case $\alpha=1/\sqrt{4.3}$), the other curves correspond to $\alpha=1/\sqrt{2}$.}
    \label{fig2}
\end{figure*}
\begin{figure*}[tp]
\includegraphics[scale=.74]{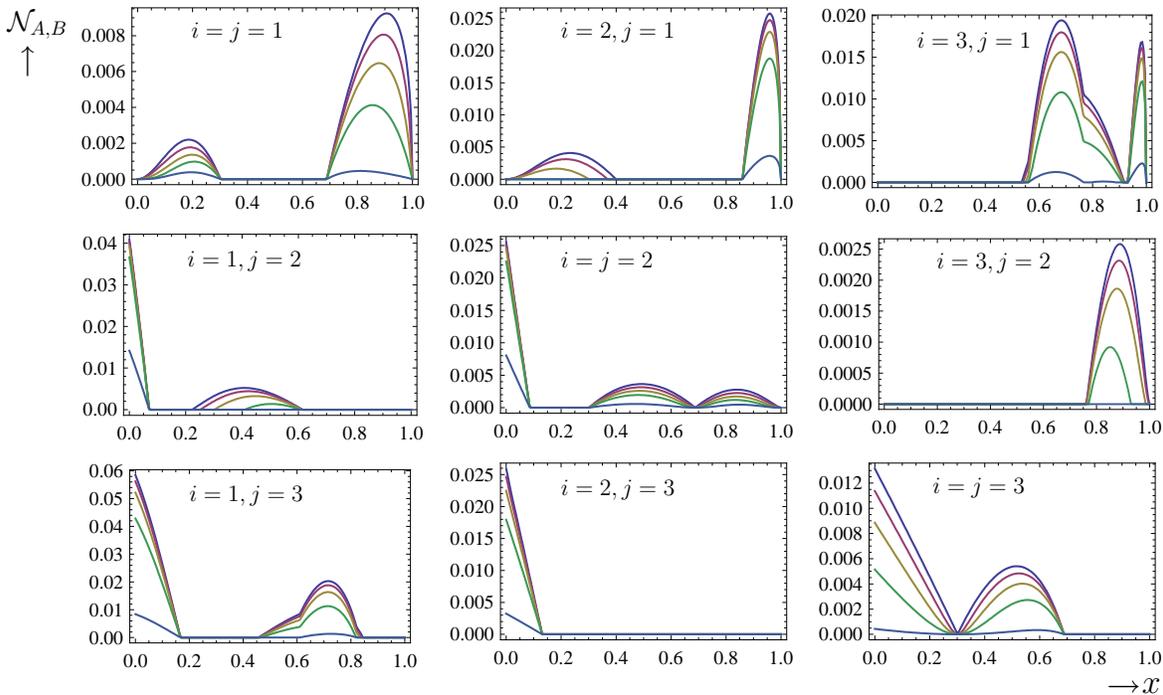}
\caption{(Color online) Entanglement of the state of Alice and Bob after weak measurement as a function of $x$, when the initial state is $\chi_2(a)$ and $\alpha=1/\sqrt{2}$. In each panel, different curves pertain to the values of $a=1$, $3/4$, $1/2$, $1/4$, and $1/50$ (from top to bottom).}
    \label{fig3}
\end{figure*}
\begin{figure*}[tbp]
      \includegraphics[scale=.76]{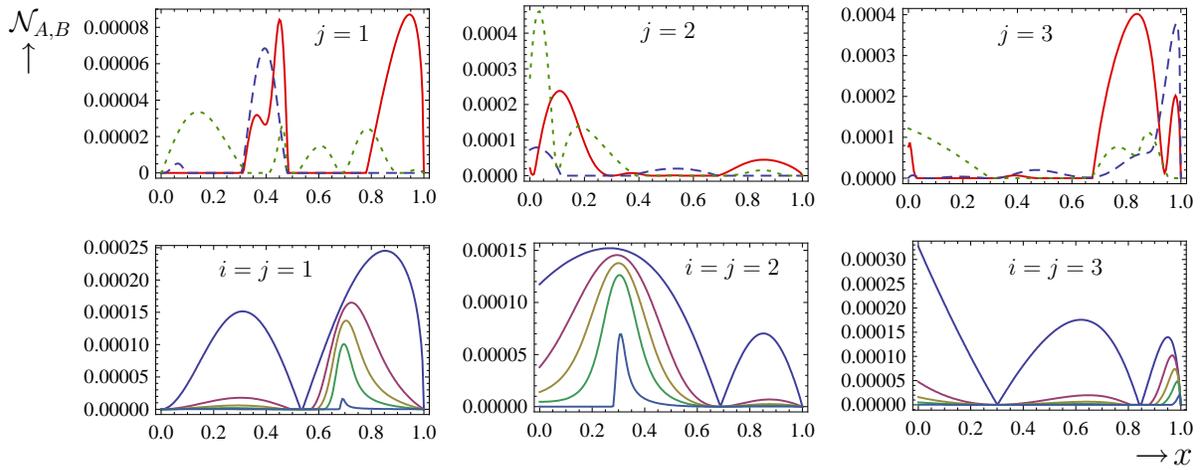}
    \caption{(Color online) Free entanglement between Alice and Bob in terms of $x$ for different outcomes when $\alpha=1/100$ and the initial states are  $\chi_1$ (up) and $\chi_2(a)$ (down, for different values of $a$ mentioned in the caption of Fig.~\ref{fig3}). For the initial state $\chi_1$, solid, dashed, and dotted curves correspond to $i=1$, $2$, and $3$, respectively. For the initial state $\chi_2(a)$, the results corresponding to the same measurement outcomes $i=j$ are depicted.}
    \label{fig4}
\end{figure*}
\begin{figure*}[tbp]
    \includegraphics[scale=.76]{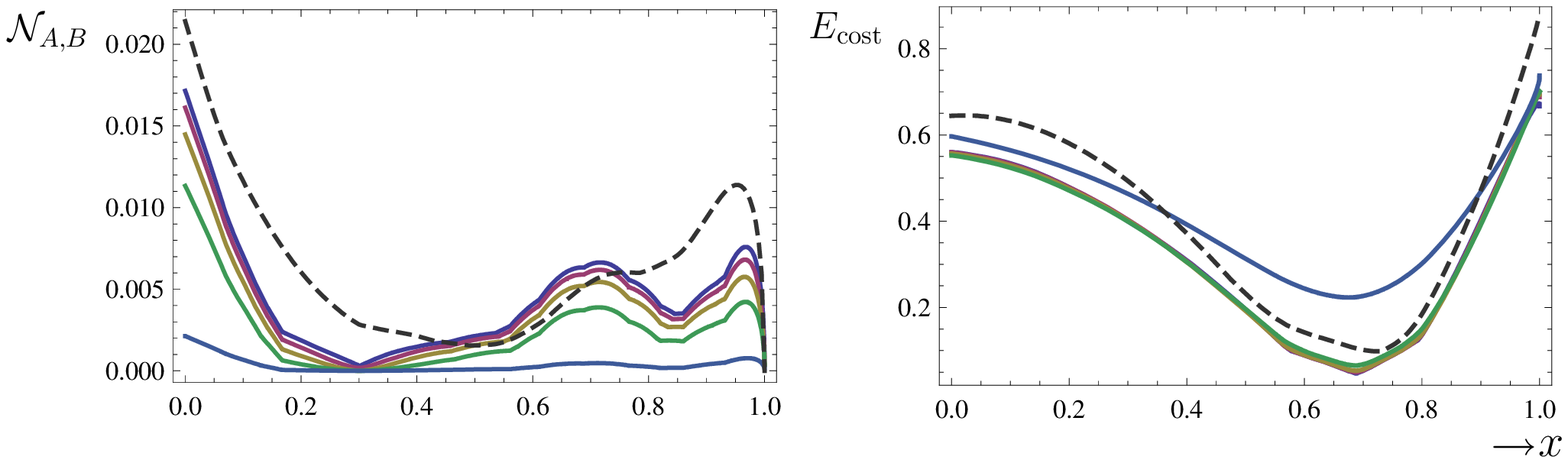}
    \caption{(Color online) The average negativity of Alice and Bob's state after measurement (left) and entanglement cost (right) vs $x$ for both initial states $\chi_1$ (dashed curves) and $\chi_2(a)$ (and different values of $a$ mentioned in the caption of Fig.~\ref{fig3}) when $\alpha=1/\sqrt{2}$. The top curve in the right panel corresponds to $\chi_2(1/50)$. It is remarkable that $\overline{{\mathcal N}}_{A,B}$ takes a local maximum for significantly weak measurements, where the entanglement cost has a local minimum.}
    \label{fig5}
\end{figure*}
and $\chi_3(b)=(2|\Phi^+\rangle\langle\Phi^+|+b\Sigma_+ +(5-b)\Sigma_-)/7$, in which $|\Phi^+\rangle=(|00\rangle+|11\rangle+|22\rangle)/\sqrt{3}$, $\Sigma_{+}=(|01\rangle\langle01|+|12\rangle\langle12|+|20\rangle\langle20|)/3$, and $\Sigma_-$
is the swap of $\Sigma_+$~\cite{act}. The state $\chi_2$ is a function of parameter $a\in[0,1]$. This state is separable when $a\in\{0,1\}$, otherwise it is bound-entangled~\cite{BEH}. The state $\chi_3$ is a function of a parameter $b\in[2,5]$. For $b\in[2,3]$ this state is separable, for $b\in(3,4]$ it is bound-entangled, and otherwise it is free-entangled~\cite{act}.

To quantify the free entanglement created via measurement, we use ``negativity" \cite{Negativity}, defined for a bipartite $d\times d'$ system $\varrho_{AB}$ (when $d<d'$) as~\cite{NNegativity}
\begin{eqnarray}
\mathcal N= (\Vert \varrho^{T_B}\Vert_1-1)/(d-1).
\end{eqnarray}
Here $\varrho^{T_B}$ denotes partial transposition with respect to the second party, which according to the Peres-Horodecki criterion~\cite{PH}, is negative when the state $\varrho_{AB}$ is free-entangled. In this case, $\mathcal N$ is positive, otherwise it vanishes. Recall that $\Vert B\Vert_1=\mathrm{Tr}[\sqrt{B^\dagger B}]$, and in a given basis $\varrho^{T_B}_{ij,kl}=\varrho_{il,kj}$.

We note that, since the positivity of the partial transposition of a state does not change with LOCC~\cite{BE}, one may need to consume some sort of nonlocality (at least indirectly) in order to transform bound entanglement into free entanglement. We share one of the states, $\chi_1$, $\chi_2(a)$, or $\chi_3(b)$, between Alice and Bob, and attach an ancillary qubit $\varrho^C$ to this system. Now, first Bob (next Alice) performs a joint weak measurement on his (her) own particle and this ancilla as follows:
\begin{eqnarray}
 M_i=\sum_{j=1}^3 \varepsilon_{j\oplus(i-1)} P_j,
\end{eqnarray}
where $i\in\{1,2,3\}$, $\oplus$ denotes modulo $3$ sum, $\varepsilon_l\in[0,1]$ are some real parameters such that $\sum_{l=1}^3\varepsilon_l^2=1$ (hereafter we rename $\varepsilon_0$ as $\varepsilon_3$), and $P_j$s are some orthogonal projectors (specified later) which satisfy $\sum_{j=1}^3 P_j=\openone$. It is thus evident that $\sum_{i=1}^3 M^\dagger_i M_i=\openone$. The result of this weak measurement can be described as
\begin{eqnarray}
{\mathcal E}_\zeta(X)=(1-\zeta){\mathcal E}_\mathrm{strong}(X)+\zeta X,\label{weakness}
\end{eqnarray}
where $\zeta=\varepsilon_1\varepsilon_2+\varepsilon_2\varepsilon_3+\varepsilon_1\varepsilon_3$, and ${\mathcal E}_\mathrm{strong}(X)=\sum_{j=1}^3 P_j X P_j$ denotes the strong or projective measurement. That is, with probability $(1-\zeta)$ the strong measurement ${\mathcal E}_\mathrm{strong}$ is applied, whereas the state does not undergo any change with probability $\zeta$. The smaller $\zeta$ is, the stronger the measurement is.

The post-measurement state of Alice, Bob, and ancilla becomes
\begin{eqnarray}
 \varrho^{ABC}_{ij}=\frac{M^{AC}_j M^{BC}_i\left(\chi^{AB}\otimes\varrho^C\right) M^{BC}_i M^{AC}_j}{\mathrm{Tr}\left[M^{AC}_j M^{BC}_i\left(\chi^{AB}\otimes\varrho^C\right) M^{BC}_i M^{AC}_j\right]},
\end{eqnarray}
where we choose
\begin{eqnarray}
&P_1\equiv\openone_{3\times2}-|\phi\rangle\langle\phi|-|\psi\rangle\langle\psi|,\nonumber\\
&P_2\equiv|\phi\rangle\langle\phi|\;,\;P_3\equiv|\psi\rangle\langle\psi|,
\end{eqnarray}
with $|\phi\rangle=\alpha|00'\rangle+\sqrt{1-\alpha^2}|11'\rangle$ and $|\psi\rangle=\sqrt{1-\alpha^2}|00'\rangle-\alpha|11'\rangle$, where $\alpha\in(0,1)$. Here $|0'\rangle$ and $|1'\rangle$ are the basis vectors of the Hilbert space of the ancillary qubit. In addition, we choose $\varepsilon_1=x$, $\varepsilon_2=\sqrt{\beta(1-x^2)}$, and $\varepsilon_3=\sqrt{(1-\beta)(1-x^2)}$, in which $x\in[0,1]$. Note that the case $\beta=1/2$ (i.e., $\varepsilon_2=\varepsilon_3$) is not of interest because in this case no entanglement is induced by measurement between the bound-entangled state and the ancilla when the outcome of measurement is $M_1$. Figure~\ref{fig1} depicts the behavior of $\zeta$ vs parameter $x$, for some fixed values of $\beta$. We shall demonstrate that our protocol generates free entanglement from bound entanglement when the measurements are generically weak.

\section{Distillation of free entanglement}

For specificity, hereon we fix the value of $\beta$ to $1/10$. Moreover, we take $\alpha=1/\sqrt{2}$, for which the entanglement content of the measurement projectors $P_2$ and $P_3$ becomes maximal. However, we also discuss the $\alpha\neq1/2$ in the sequel as well. Furthermore, we initialize the ancilla qubit in the state $(|0'\rangle+|1'\rangle)/\sqrt{2}$.

Suppose that initially Alice and Bob share the state $\chi_1$. Figure~\ref{fig2} shows the variation of negativity of their post-measurement state, $\varrho_{ij}^{AB}=\mathrm{Tr}_C[\varrho_{ij}^{ABC}]$, in terms of $x$, for different outcomes of measurements. It is remarkable that in this case, all $x\in(0,1/4]$ yield the free entanglement with certainty.

\begin{figure}[bp]
    \includegraphics[scale=.77]{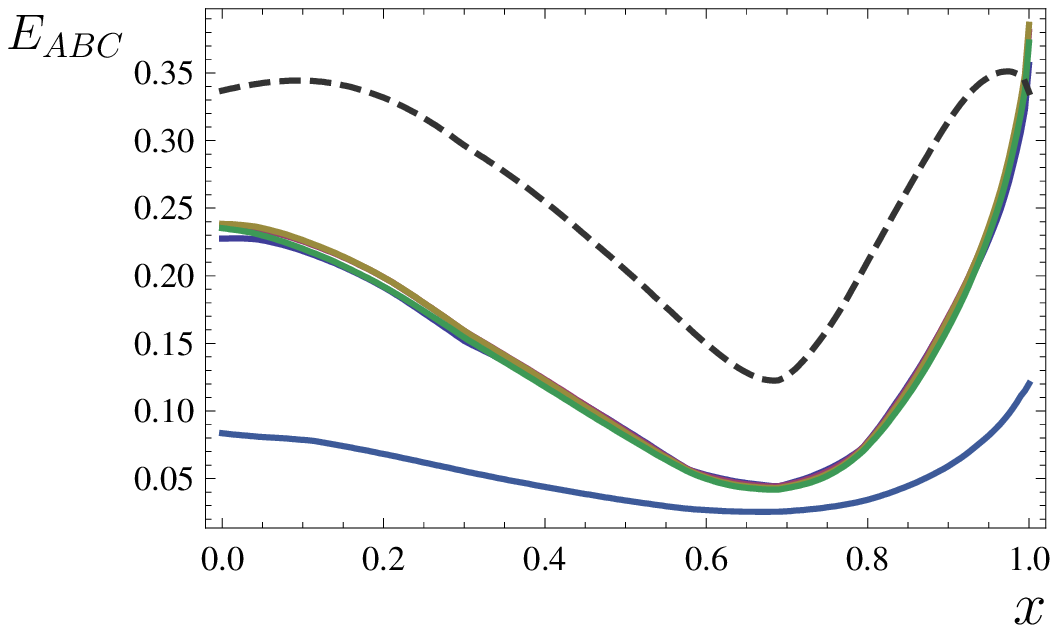}
    \caption{(Color online) Genuine tripartite entanglement, $E_{ABC}=\overline{{\mathcal N}}^2_{AB,C}-(\overline{{\mathcal N}}^2_{A,C}+\overline{{\mathcal N}}^2_{B,C})$~\cite{Monogamy}, as a function of $x$, for initial states $\chi_1$ (dashed curve) and $\chi_2(a)$ (for the values of $a$ as in the previous figures) when $\alpha=1/\sqrt{2}$. The lowest curve corresponds to $\chi_2(1/50)$.}
    \label{fig6}
\end{figure}

If Alice and Bob begin with the state $\chi_2(a)$, negativity of the post-measurement state behaves as in Fig.~\ref{fig3}. This figure demonstrates that: (i) in all outcomes, there exist intervals of $x$ in which transition from bound entanglement to free entanglement occurs. In particular, when $x\in(0,0.05]$, one can assure that the transition takes place with a high probability; that is, seven out of nine total cases show nonvanishing amounts of free entanglement. (ii) For all allowed values of $a\neq0$ this transition occurs (for some values of $x$). (iii) For $a$s not too close to zero, the maximum amount of free entanglement is created when the measurement outcomes $M_1$ and $M_3$ are obtained for Bob and Alice, respectively.

A similar analysis shows that when $\chi_3(b)$ is the initial state, after the measurement, free entanglement is generated between Alice and Bob for all $b\in[2,4]$ with nonzero probability---plots not presented here. The overall behavior of ${\mathcal N}_{A,B}$, for each measurement outcome is also akin to that of $\chi_2(a)$, with the difference that in this case, even when the measurement outcomes are the same ($i=j$), the points at which negativity vanishes depend on both $b$ and $x$.

If $\alpha$ approaches $0$ or $1$, the maximum value of $\mathcal{N}_{A,B}$ may increase or decrease depending on the outcomes and the value of $a$ and $b$ for the initial states $\chi_2(a)$ and $\chi_3(b)$. For example, in the case of the initial state $\chi_1$ and the outcome $M_2^{AC} M_3^{BC}$, $\mathcal{N}_{A,B}$ takes its maximum value at $\alpha\approx 1/\sqrt{4.3}$ (see the dot-dashed curve in the middle panel of Fig.~\ref{fig2}). The interval of $x$ in which the transition occurs may also increase or decrease for all initial states, but here the point is that for all $\alpha\neq 0,1$ one can distill free entanglement with nonzero probability (for the initial state $\chi_3(b)$, this is correct for $b=4$). For example, we showed the results corresponding to $\alpha=1/100$ in Fig.~\ref{fig4}.

To get further insight on the cost of generating free entanglement between Alice and Bob, we investigate the non-locality content of the applied (weak) measurements. Note that the pre-measurement state of Bob and ancilla is a product state. Thus, the average entanglement of their post-measurement state can be considered as a lower bound on the nonlocality that this measurement contains. A similar argument is also applicable to the measurement performed on the Alice-ancilla state in the next step since this state before measurement is separable---because it can be checked that its negativity is zero and since this state is $3\times2$, positivity of its partial transposed implies its separability. We remark that the method proposed in Ref.~\cite{EntCost} to identify nonlocal cost of a measurement does not seem suitable to apply in our scenario, because in that method the pre-measurement state is maximally mixed, and the method uses a clever trick to remove the effect of the initial state on the nonlocality content of a measurement. Taking all these points into account, an estimate for nonlocality on the measurements in our scheme is as follows:
\begin{eqnarray}
\hskip-.5mm M_\mathrm{cost}=\sum_{j=1}^3\left[p_{B,C}(j)\mathcal{N}_{B,C}(j)+\sum_{k=1}^3 p_{A,C}(j,k)\mathcal{N}_{A,C}(j,k)\right]\hskip-1.2mm,\nonumber
\end{eqnarray}
where $p_{B,C}(j)$ [$p_{A,C}(j,k)$] in the first (second) term denotes the probability of obtaining the result(s) $j$ ($j$ and $k$) after a joint measurement over party $B$ and ancilla $C$ ($A$ and ancilla $C$), and $\mathcal{N}_{B,C}(j)$ [$\mathcal{N}_{A,C}(j,k)$] is the negativity of the post-measurement state of the same parties, corresponding to the same result(s). We consider $E_\mathrm{cost}\equiv M_\mathrm{cost}-\overline{{\mathcal N}}_{A,B}$ as the entanglement cost of our distillation scenario (depicted in the right panel of Fig.~\ref{fig5}). Here, $\overline{{\mathcal N}}_{A,B}$ denotes the average of free entanglement generated between Alice and Bob after the measurements (see the left panel of Fig.~\ref{fig5}). The curves shown in the right panel of Fig.~\ref{fig5} present the behavior of $E_\mathrm{cost}$ versus $x$ for different initial states $\chi_1$ and $\chi_2(a)$. It should be remarked that, in our protocol, measurements indeed can generate entanglement in the $(AB)C$, $A(BC)$, and $B(AC)$ bipartitions as well as genuine tripartite entanglement (Fig.~\ref{fig6}). Nevertheless, since here we are only interested in the generated entanglement between $A$ and $B$, in our analysis we only subtract this entanglement from the measurement cost in order to obtain a lower bound for the entanglement cost of our scenario. One should also note that the existence of all these correlations naturally restricts the amount of free entanglement generated between Alice and Bob after measurements. Figures~\ref{fig1} and \ref{fig5} imply that if Alice and Bob choose the measurement strength from the significantly weak ranges, e.g., $x\in(0.65,0.75)$, they can distill relatively high amounts of free entanglement with relatively low cost. Having said all this, to give a better estimate of how much nonlocality is needed in our scenario, a more detailed analysis in which all sorts of entanglement (bipartite and tripartite) are taken into account is needed. For our purposes, however, the given analysis suffices.

A final remark is in order here. It might be argued that our protocol transfers the initial bound entanglement to the other partitions. However, it is straightforward to show that in the range of parameters where distillation of free entanglement between $A$ and $B$ is successful there does not exist any further bipartite bound entanglement in the total system. Accordingly, our protocol does not distribute bound entanglement within the system.

\section{Summary}

Here we have proposed a controllable scheme for distilling free entanglement from bipartite bound-entangled states. Unlike previous entanglement distillation protocols, our protocol employs an ancillary qubit, and is based on weak measurements, obviating the need to share any free-entangled state between the parties. In this sense, our protocol uses a non-distillable entangled state and transforms it into a ``useful" type without having to consume other useful states. Rather, the entanglement is unlocked through the very measurement process. Therefore, in order to analyze the cost of our protocol, we have compared how much entanglement needs to be invested in the measurement process, and as a result how much entanglement can be obtained. This argument has implied that our protocol can generate useful entanglement with a set of measurements which do not need much entanglement to be realized.

There may still exist projective or strong measurements which can also transform bound to free entanglement. Our scheme can be a suitable alternative in situations, e.g., where physical realization of such strong measurements is difficult. Although we have illustrated our protocol with specific $3\times 3$ bound-entangled states, generalization to higher dimensional and multipartite systems seems straightforward.\\[1cm]

\begin{acknowledgments}
We appreciate valuable discussions with V. Karimipour. ATR acknowledges partial financial support by Sharif University of Technology's Office of Vice-President for Research.
\end{acknowledgments}


\end{document}